\documentclass[12pt,fleqn]{article}
\usepackage{psfig}
\begin{document}
\begin{center}
{\Large{\bf Chiral symmetry in electromagnetically induced resonant two hadron
production}}
\end{center}

\vspace{.7cm}

{\large{ E. Oset$^ {1,2}$, A. Ramos$^ {3}$, E. Marco$^ {1,2}$, 
J. C. Nacher$^{1,2}$, H. Toki$^{1}$}}

\vspace{0.8cm}

{\it $^1$ Research Center for Nuclear Physics (RCNP), Osaka University, Ibaraki,
 Osaka 567-0047, Japan.}

{\it $^2$  Departamento de F\'{\i}sica Te\'orica and IFIC, 
Centro Mixto Universidad de Valencia-CSIC
46100 Burjassot (Valencia), Spain.}

{\it $^3$ Departament d'Estructura i Constituents de la Materia, Universitat de
Barcelona, Diagonal 647, 08028 Barcelona, Spain.}

\vspace{0.6cm}

\begin{abstract}
{\small{We report on recent progress on the 
chiral unitary approach, analogous to the effective range expansion
in Quantum Mechanics, which is shown to have a much larger convergence radius
than ordinary chiral perturbation theory, allowing one to reproduce data for
meson meson interaction up to 1.2 GeV. Applications to physical processes so
far unsuited for a standard chiral perturbative approach are presented, in which
two hadrons, either two mesons or a meson and a baryon, in a resonant state are
produced in photoproduction processes.}} 
\end{abstract}
\vspace{.5cm}

\vspace{0.5cm}
\section{Chiral Unitary Approach}

Chiral perturbation theory ($\chi PT$) has proved to be a  very suitable
instrument to implement the basic dynamics and symmetries of the meson meson
and meson baryon interaction \cite{gasser} at low energies. The essence of the
perturbative technique, however, precludes the possibility of tackling problems
where resonances appear, hence limiting tremendously the realm of
applicability. The method that we expose naturally leads to low lying
resonances and allows one to face many problems so far intractable within
$\chi PT$.

The method incorporates new elements: 1) Unitarity is implemented exactly;
2) It can deal with coupled channels allowed with pairs of particles from the 
octets of stable
pseudoscalar mesons and ($\frac{1}{2}^+$) baryons; 3) A chiral
expansion in powers of the external four-momentum of the lightests 
pseudoscalars
is done for
Re $T^{-1}$, instead of the $T$ matrix itself which is the case in standard $\chi
PT$.

We sketch here the steps involved in this expansion for the meson meson
interaction. One starts from a $K$ matrix approach in coupled channels where
unitarity is automatically fulfilled and writes 
\begin{equation}
T^{-1} = K^{-1} - i\,\sigma ,
\end{equation} 
where $T$ is the scattering matrix, $K$ a real matrix in the physical 
region
and $\sigma$
is a diagonal matrix which measures the phase-space available for the intermediate
states
\begin{equation} 
\sigma_{nn}(s) = - \frac{k_n}{8\pi\sqrt{s}}\,\theta\left(s - (m_{1n} + m_{2n})^2\right),
\end{equation}
where $k_n$ is the on shell CM momentum of the meson in the intermediate state
$n$ and $m_{1n}$, $m_{2n}$ are the masses of the two mesons in the state $n$. The meson
 meson states considered here are $K\bar{K}$, $\pi\pi$, $\pi\eta$, $\eta\eta$,
 $\pi K$, $\pi\bar{K}$, $\eta K$, $\eta\bar{K}$. Since $K$ is real, from eq. (1)
 one sees that $K^{-1}$ = Re $T^{-1}$. In non-relativistic
 Quantum Mechanics, in the scattering
 of a particle from a potential, it is possible to expand
 $K^{-1}$ in powers of the momentum
 of the particle at low energies as follows (in the s-wave for simplicity)
\begin{equation} 
\hbox{Re}\,T^{-1}\equiv K^{-1} = \sigma\cdot ctg\delta\, \propto -\frac{1}{a} 
+ \frac{1}{2}r_0 k^2 ,
\end{equation}
with $k$ the particle momentum, $a$ the scattering length and $r_0$ the effective
range.

The ordinary  $\chi$PT expansion up to $O(p^4)$ is given by \cite{gasser}
\begin{equation}
T = T_2 + T_4 ,
\end{equation}
where $T_2$, which is $O(p^2)$, is obtained from the lowest order chiral
Lagrangian, $L^{(2)}$, whereas $T_4$ contains one loop diagrams in the s, t, u
channels, constructed from the lowest order Lagrangian, tadpoles and the
finite contribution from the 
tree level diagrams of the $L^{(4)}$ Lagrangian. This last contribution, 
after a suitable renormalization, is just a polynomial, $T^{(p)}$.
Our $T$ matrix, starting from eq. (1) is given by
\begin{equation}
T = [\hbox{Re}\, T^{-1} - i\,\sigma]^{-1}
\equiv T_2 \,[T_2 \,\hbox{Re}\, T^{-1}\, T_2 - i\,T_2\,\sigma\, T_2]^{-1}\, T_2 ,
\end{equation}
where, in the last step, we have multiplied by $T_2 T_2^{-1}$ on the left and 
$T_2^{-1} T_2$ on the
right for technical reasons. But using standard $\chi$PT we obtain the following
expansion up to order $O(p^4)$,
\begin{equation}
T_2 \,Re T^{-1}\, T_2 = T_2 - \hbox{Re}\, T_4 ...
\end{equation}
and hence, recalling that $\hbox{Im}\, T_4 = T_2\,\sigma \,T_2$, one obtains
\begin{equation}
T = T_2\,[ T_2 - T_4]^{-1}\, T_2,
\end{equation}
which is the coupled channel generalization of the inverse amplitude method of
\cite{dob}.

        Once this point is reached one has several options to proceed:

a) A full calculation of $T_4$ within the same renormalization scheme as in
$\chi PT$ can be done. The eight $L_i$ coefficients from $L^{(4)}$ are then fitted
to the existing meson meson data on phase shifts and inelasticities up to 1.2 GeV, where
4 meson states are still unimportant. This procedure has been carried out in
\cite{dob,gue}. The resulting $L_i$ parameters are compatible with those used in $\chi PT$. 
At low energies the $O(p^4)$ expansion for $T$ of eq. (7) is identical to that
in $\chi PT$. However, at higher energies the nonperturbative structure of eq. (7),
 which implements unitarity exactly, allows one to extend the information
 contained in the chiral Lagrangians to much higher energy than in ordinary 
$\chi$ PT, which is up to about $\sqrt{s}\simeq 400 $ MeV. Indeed it
reproduces the resonances present in the L = 0, 1 partial waves.

\vskip .2cm

b) A technically simpler and equally successful additional approximation
 is generated by ignoring the crossed channel loops and
tadpoles and reabsorbing them in the $L_i$
coefficients given the weak structure of these terms in the physical region.
The fit to the data with the new $\hat{L}_i$ coefficients reproduces the whole meson
meson sector, with the position, widths and partial decay widths of the
$f_0(980)$, $a_0(980)$, $\kappa(900)$, $\rho(770)$, $K^\ast(900)$ resonances in good
agreement with experiment \cite{oller1}. A cut off regularization is used in \cite{oller1} for the
loops in the s-channel. By taking the loop function with two intermediate
mesons
\begin{equation}
G_{nn}(s) = i\int\frac{d^4 q}{(2\pi)^4}\, \frac{1}{q^2 - m_{1n}^2 + i\epsilon}
\, \frac{1}{(P-q)^2 - m_{2n}^2 + i\epsilon},
\end{equation}
where $P$ is the total meson meson momentum, one immediately notices that
\begin{equation}
\hbox{Im}\, G_{nn}(s) = \sigma_{nn}.
\end{equation}
Hence, we can write
\begin{equation}
\hbox{Re}\, T_4 = T_2\, \hbox{Re}\, G\, T_2 + T_4^{(p)},
\end{equation}
where $\hbox{Re}\, G$ depends on the cut off chosen for $|\vec{q}|$. This means that the
$\hat{L}_i$ coefficients of $T_4^{(p)}$ depend on the cut off choice, much as the
$L_i$ coefficients in $\chi PT$ depend upon the regularization scale.

\vskip .2cm

c) For the L = 0 sector (also in L = 0, S = $-1$ in the meson baryon interaction)
a further technical simplification is possible. In these cases it is possible
to choose  the cut off such that, given the relation between $\hbox{Re}\, G$
and $T_4^{(p)}$, this latter term is very well approximated by 
$\hbox{Re} T_4= T_2 \,\hbox{Re}\, G\, T_2$. This is impossible in those cases
because of the predominant role played by the unitarization of the lowest
order $\chi PT$ amplitude, which by itself leads to the low lying resonances,
 and because other genuine QCD resonances appear at higher energies.

 In such a case eq. (5) becomes 
\begin{equation}
T = T_2 \,[T_2 - T_2\, G \,T_2]^{-1} \,T_2 = [1 - T_2 \,G]^{-1}\, T_2,
\end{equation}
or, equivalently,
\begin{equation}
T = T_2   + T_2 \,G \,T,
\end{equation}
which is a Bethe-Salpeter equation with $T_2$ and $T$ factorized on shell outside
the loop integral, with $T_2$ playing the role of the potential. This option has
proved to be successful in the L = 0 meson meson sector in \cite{oller2} and in the
L = 0, S = $-1$ meson baryon sector in \cite{osetra}.

        In the meson baryon sector with S = 0, given the disparity of the
        masses in the coupled channels $\pi N$, $\eta N$, $K\Sigma$,
        $K\Lambda$,
        the simple ``one cut off approach'' is not possible. In \cite{kaiser} higher
        order Lagrangians are introduced while in \cite{par} different subtraction
        constants (or equivalently different cut offs) in G are incorporated
        in each of the former channels leading in both cases to acceptable
        solutions when compared with the data.

\section{Application to the photoproduction of meson baryon pairs in resonant
states}

As quoted above, a good description of the $K^-p$ and coupled channel interaction 
is obtained in terms of the lowest order Lagrangians and the Bethe Salpeter
equation with a single cut off. One of the interesting features of the approach
is the dynamical generation of the $\Lambda(1405)$ resonance just below the
$K^-p$ threshold. The threshold behavior of the $K^-p$ amplitude is thus
very much tied to the properties of this resonance. Modifications of these
properties in a nuclear medium can substantially alter the $K^-p$ and $K^-$
nucleus interaction and experiments looking for these properties are most welcome. Some 
electromagnetic reactions appear well suited for these studies.
Application of the chiral unitary approach to the
$K^-p\rightarrow\gamma\Lambda$, $\gamma\Sigma^0$ reactions at threshold has
been carried out in \cite{lee} and a fair agreement with experiment is found. In
particular one sees there that the coupled channels are essential to get a good
description of the data, increasing the $K^-p\rightarrow\gamma\Sigma^0$ rate
by about a factor 16 with respect to the Born approximation.

        In a recent paper \cite{nac1} we propose the $\gamma p\rightarrow
        K^+\Lambda(1405)$ reaction as a means to study the properties of the
        resonance, together with the $\gamma A\rightarrow
        K^+\Lambda(1405) A'$ reaction to see the modification of its properties
        in nuclei. The resonance $\Lambda(1405)$ is seen in its decay products
        in the $\pi\Sigma$ channel, but as shown in \cite{nac1} the sum of the cross
        sections for $\pi^0\Sigma^0$, $\pi^+\Sigma^-$, $\pi^-\Sigma^+$
        production has the shape of the resonance $\Lambda(1405)$ in the I = 0
        channel. Hence, the detection of the $K^+$ in the elementary reaction,
         looking at $d\sigma/dM_I$ ($M_I$ the invariant mass of the meson
         baryon system which can be induced from the $K^+$ momentum), is
         sufficient to get a clear $\Lambda(1045)$ signal. In nuclear targets
         Fermi motion blurs this simple procedure (just detecting the $K^+$), but the
         resonance properties can be reconstructed by observing the decay
         products in the $\pi\Sigma$ channel. In fig. 1 we show the cross
         sections predicted for the $\gamma p\rightarrow K^+ \Lambda(1405)$
         reaction looking at $K^+\pi^0\Sigma^0$, $K^+$ $all$ and $K^+ \Lambda(1405)$
         (alone). All of them have approximately the same shape and strength
         given
         the fact that the I = 1 contribution is rather small.
\vspace{0.3cm}
\begin{figure}[h]
\centerline{\protect
\hbox{
\psfig{file=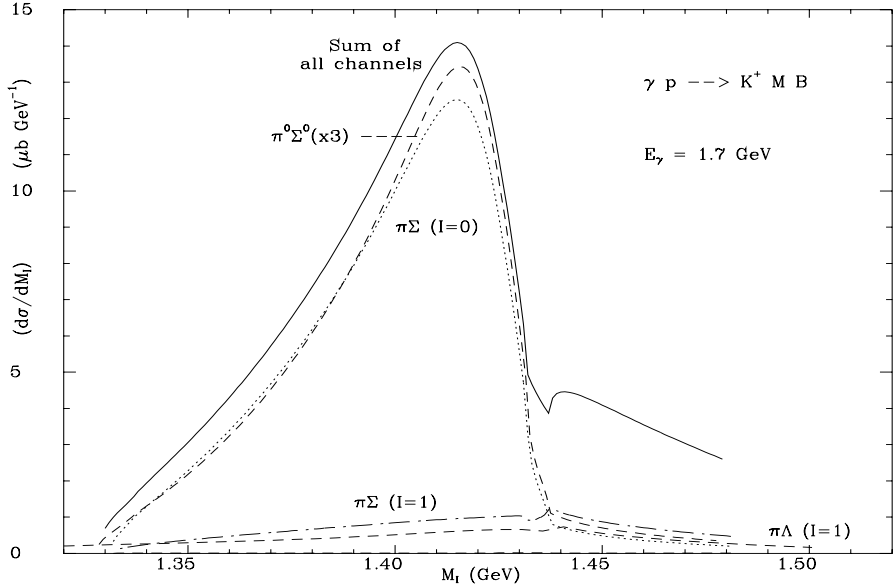,height=8cm,width=13cm,angle=-90}}}
\vspace{0.5cm}
\caption{Cross section for $\gamma p\rightarrow K^+ X$ with $X=all$,
$\pi^0\Sigma^0$, $\Lambda$(1405).}
\end{figure}
\vspace{0.5cm}


        The energy chosen for the photon is $E_\gamma$ = 1.7 GeV which makes it
        suitable of experimentation at SPring8/RCNP, where the experiment is
        planned \cite{nakano}, and TJNAF.
        
        One variant of this reaction is its time reversal $K^- p
\rightarrow\Lambda(1405)\gamma$. This reaction, for a $K^-$ momentum
in the 300 to 500 MeV/c range, shows clearly the $\Lambda(1405)$ resonant production
\cite{nac2} and
has the advantage that the analogous reaction in nuclei still allows the
observation of the $\Lambda(1405)$ resonance with the mere detection of the
photon, the Fermi motion effects being far more moderate than in the case of
the $\gamma A\rightarrow K^+\Lambda(1405)X$ reaction which requires larger
photon momenta and induces a broad distribution of $M_I$ for a given $K^+$
 momentum.

 One of the interesting developments around these lines is the interaction of
 the $K^-$ with the nuclei, with its relationship to problems like
 $K^-$ atoms or the possible condensation of $K^-$ in neutron stars. The
 problem has been looked at from the chiral perspective by evaluating Pauli
 blocking effects on the nucleons of the intermediate $\bar{K}N$ states
 \cite{koch,waas}.
  These effects lead to a $K^-$ self-energy in nuclei which is attractive
  already at very low densities, as a consequence of pushing the resonance at
  energies above $K^-p$ threshold. However, more recent investigations
  considering the $\bar{K}$ self-energy in a self-consistent way \cite{lutz} lead to
  quite different results since the resonance barely changes its position. Yet
  one still gets an attractive self-energy which is demanded by the $K^-$ atom
  data \cite{batty}. A step forward in this direction is given in \cite{ramos}, where in
  addition to the $K^-$ self-energy in the medium, one also renormalizes the
  pions and takes into account the different binding of N, $\Sigma$ and
  $\Lambda$ in nuclei. Preliminary results from \cite{oku} indicate that the 
  $K^-$ self-energy obtained in \cite{ramos} can lead to a  good microscopical
  description of present data on $K^-$ atoms, hence providing an accurate tool
  to study the properties of $K^-$ at higher densities and the eventual
  condensation in neutron stars.
 \vspace{0.2cm}

\section{Photoproduction of resonant two meson states}

    Another application which can be done using the same reaction is the
 photoproduction of resonant two meson states. Particularly the $f_0$(980) and
 $a_0$(980) resonances. These states appear in $L=0$ in isospin zero and one
 respectively. The scalar sector of the meson is very controversial and the
 chiral unitary theory has allowed one to bring a new perspective on these
 states. Concretely the $a_0$(980) is a resonant state of two mesons, 
 mostly $K\bar{K}$
 coming from the interaction of the mesons, $\pi\eta$ and $K\bar{K}$, in the coupled
 channel unitary approach, while the $f_0$(980) state is still mostly a resonant
 state of $K\bar{K}$ and $\pi\pi$, but it has also a small admixture of a
 genuine, $q\bar{q}$,
 state, a singlet of SU(3), which appears around 1 GeV in a study of the meson
 meson interaction in which the lowest order of $\chi PT$ is implemented together with
 the explicit exchange of preexisting resonances which provide the contribution
 to the higher order terms of the chiral Lagrangians \cite{ollernew}. 
   In the present case the reaction suggested is \cite{marco}
 
\begin{equation}
\gamma p\rightarrow p M
\end{equation}
where $M$ is either of the resonances $a_0$(980) or $f$(980). In practice the meson M
 will decay into two mesons , $\pi\pi$ or $K\bar{K}$ in the case of the 
 $f_0$(980)  or $K\bar{K}$ , $\pi\eta$ in the case of the $a_0$(980).                                       
 \begin{figure}
\centerline{
\hbox{
\centerline{\protect\hbox{
\psfig{file=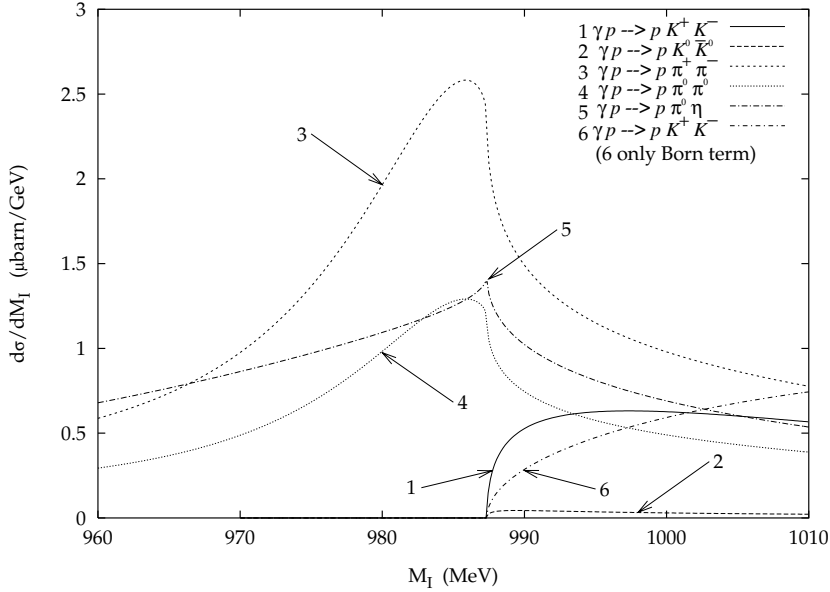,width=0.8\textwidth,silent=,angle=-90}}}
}}
\caption{Results for the photoproduction cross section on protons
as a function of the invariant
mass of the meson-meson system.}
\end{figure}                                        

In Fig.~2 we show the results for the 5 channels considered. We observe
clear peaks for $\pi^+ \pi^-$, $\pi^0 \pi^0$ and $\pi^0 \eta$
production around 980 MeV. The
peaks in $\pi^+ \pi^-$ and $\pi^0 \pi^0$ clearly correspond to the
formation of the $f_0(980)$ resonance, while the one in
$\pi^0 \eta$ corresponds to the formation of the
$a_0(980)$. The $\pi^0 \pi^0$ cross section is $\frac{1}{2}$
of the $\pi^+ \pi^-$ one due to the symmetry factor . 
The $K^+ K^-$ and $K^0 \bar{K}^0$ production cross section appears
at energies higher than that of the resonances and hence do not show
the resonance structure. Yet, final state interaction is very important
and increases appreciably the $K^+ K^-$ production cross section
for values close to threshold with respect to the Born approximation.

It is also interesting to see the shapes
of the resonances which differ appreciably from a Breit-Wigner, due
to the opening of the $K \bar{K}$ channel just above the
resonance \cite{flatte}.

We would like to stress here that the invariant
mass distributions for resonance excitation into the various
pseudoscalar channels depicted in Fig.~2 are theoretical predictions
of a chiral unitary model, in this case the one of [5],
where only one parameter was fitted to reproduce all the data of
the meson-meson interaction in the scalar sector.

A small variant of this reaction would be the 
$\gamma p \rightarrow n M \bar{M}$. In this case the $M \bar{M}$
system has charge $+1$ and hence $I=0$ is excluded, hence,
one isolates the $a_0$ production.

It is interesting to notice that the $f_0$(980) resonance shows up as a peak
in the reaction. This is in contrast to the cross section for
$\pi \pi \rightarrow \pi \pi$ in $I=0$ which exhibits a minimum at the
$f_0$ energy because of the interference between the $f_0$
contribution and the $\sigma(500)$ broad resonance. 

However, we should bear in mind that we
have plotted there the contribution of the $f_0$ resonance alone. The tree
level contact term and Bremsstrahlung diagrams, plus other contributions
which would produce a background, are not considered there. We estimate
the background from the experimental cross section for
$\gamma p \rightarrow p \pi^+ \pi^-$ of \cite{cola}, which is
around 45 $\mu b$ at $E_{\gamma} = 1.7$ GeV. This provides a background of around 55 $\mu b /$GeV while
the resonant peak has about 2.5 $\mu b/$GeV strength. This gives a ratio
of 5\% signal to background assuming that the
background is mostly real versus an imaginary contribution
from the resonance and hence there would be no interference.
The situation with the $\pi^0 \pi^0$ channel should be better
because the $\gamma p \rightarrow \pi^0 \pi^0 p$ cross section
is about eight times smaller than the one for
$\gamma p \rightarrow \pi^+ \pi^- p$ \cite{Zabrodin,Harter}.
Considering that the resonant signal now is a factor two smaller than the
$\gamma p \rightarrow \pi^+ \pi^- p$ cross section, this would give
a ratio of signal to background of 20\%, which should be more clearly
visible in the experiment. The same or even better ratios than in
the $\pi^0 \pi^0$ case are expected
for $\pi^0 \eta$ production in the $a_0$ channel, since estimates
of the background along the lines of present models for
$\pi^0 \pi^0$ production \cite{Tejedor,Ochi} would provide a
cross section smaller than for $\pi^0 \pi^0$ production. 

However, there is a distinct feature about the $f_0$ resonance which
makes its contribution, in principle, bigger than the estimates
given above. Indeed, the $f_0$ is approximately a Breit-Wigner resonance
with an extra phase of $e^{i \pi/2}$. This means that the real part
has a peak while the imaginary part changes sign around the
resonance energy. This means that assuming the background basically real,
there would be an interference with the $f_0$ resonance which would
lead to an increase of about 50\% over the background, or a decrease
by about 40\% (depending on the relative sign) for the $\pi^+ \pi^-$
case and larger effects for the $\pi^0 \pi^0$ case. This is of course
assuming weak dependence on momenta and
spin of the background amplitudes. In any case, due to the particular
feature of the $f_0$ resonance discussed above, it is quite
reasonable to expect bigger signals than the estimates based on a
pure incoherent sum of cross sections.

Certainly it is possible to obtain better ratios if one looks at angular
correlations. If one looks in a frame where the two mesons are
in their CM, the Bremsstrahlung pieces (both from the squared of
the Bremsstrahlung term as well as from interference with
s-wave terms) have a $\sin^2 \theta$ dependence,
with $\theta$ the angle between the meson and the photon.
Other terms from \cite{Tejedor,Ochi} exhibit equally strong
angular dependence, for what extraction of the angle independent part
of the cross section would 
be an interesting exercise which would select the part
of the cross section to which the resonant contribution
obtained here belongs to.

\section{ Summary}

  We have reported on the unitary approach to meson meson and meson baryon
  interaction using chiral Lagrangians, which has proved to be an efficient
  method to extend the information contained in these Lagrangians to higher
  energies where $\chi PT$ cannot be used. This new approach has opened the
  doors to the investigation of many problems so far intractable with $\chi PT$
  and a few examples have been reported here. We have applied these techniques
  to the problem of photoproduction of scalar mesons $f_0$(980), $a_0$(980) and
  the photoproduction of the $\Lambda(1405)$, a resonant state of meson baryon
  in the $S=-1$ sector and have found signals which are well within measuring
  range in present facilities. The experimental implementation of these
  experiments contrasted with the theoretical predictions will contribute with new
  tests of these emerging pictures implementing chiral symmetry and unitarity,
  which for the moment represent the most practical approach to QCD at low
  energies.
\vspace{0.2cm}

\section*{Acknowledgments.}
\vspace{0.1cm}
        
        We are thankful to the COE Professorship program of Monbusho which enabled
        E. O. to stay at RCNP where part of the work reported here has been
        done. E. M. and J. C. N. would like to acknowledge the hospitality
        of the RCNP of the Osaka University and support from the Ministerio de
        Educacion y Cultura. This work is partly supported by DGICYT, contract
        number PB 96-0753.

\end{document}